\documentclass{llncs}

\usepackage{array}
\usepackage{pstricks}
\usepackage{graphicx}
\usepackage{amsfonts}
\usepackage{algorithm,algorithmic}

\newcommand{\tb}{\textbackslash}

\title{Multiple Tree for Partially Observable Monte-Carlo Tree Search}

\author{David Auger}

\institute{tao, LRI, Orsay\\%
\& INRIA Saclay}

\begin{document}
\maketitle

\begin{abstract}
   We propose an algorithm for computing approximate Nash equilibria of partially observable games using  Monte-Carlo tree search based on recent bandit methods. We obtain experimental results for the game of phantom tic-tac-toe, showing that strong strategies can be efficiently computed by our algorithm.
\end{abstract}

\section{Introduction}

In this paper, we introduce a method for computing Nash equilibria in partially observable games with large state-space.
Partially observable games - also called games with incomplete information - are games where players know the rules but cannot fully see
the actions of other players and the real state of the game, e.g. card games. Among these games, a classical testbed for computer algorithms are phantom games, 
the most well known being Kriegspiel \cite{li94}, and computer scientists often consider phantom-go \cite{cazenave06}. We here focus on a simpler game, namely phantom tic-tac-toe, which is still unsolved;
our algorithm is nonetheless a generic tool for partially observable games. 

The game of phantom tic-tac-toe (a.k.a. noughts and crosses) is played on a $3 \times 3$ grid. The players take turns, respectively marking with ``X'' and ``O'' the squares of the grid, and the first player to obtain three of his marks in an horizontal, vertical or diagonal row wins the game. The difference between the standard and the phantom game is that in the latter, players do not see where their opponent plays. 
If they try to play in an ``illegal'' square,
then they are informed of this fact and must play somewhere else. Playing such an illegal 
move is never harmful since it brings information about the real state of the game, and good strategies will use this.

The game of phantom tic-tac-toe, as well as numerous other games like chess, go or poker, can be modelled in the so called {\it extensive form},
which is given by a tree where nodes correspond to the different positions of the game, and arcs to the decisions of players
(see e.g. \cite{fudenberg91}). In the partial observation case, we must add to this framework {\it information sets}
grouping  nodes that a player cannot distinguish.

When the game has full observability, Monte-Carlo tree search \cite{coulom06} (MCTS for short) 
is known as a very efficient tool for computing strong strategies. 
Let us describe briefly how such an algorithm works. The algorithm grows a subtree $T_1$ of the whole tree of the game $T$.
For each new round, the algorithm simulates a single play by moving in the tree $T$ down from the root. The tree $T$ does not have to be stored, but is implicitly given by the rules of the game. For each node of $T$ where some player has to make a decision, two case may happen:
\begin{itemize}
  \item either the node is in $T_1$, and then a decision is made according to information stored in this node;
  \item either the node is not in  $T_1$, and a move is randomly chosen, generally with uniform probability.
\end{itemize}

When the simulation ends, either by a player's victory or a draw, the first encountered node of $T$ which is not
in $T_1$ is added to $T_1$, and in this node up to the root, informations concerning the last simulation are processed (usually,
the number of simulations and victories where these nodes were encountered).

The policy used to choose in $T_1$ between different actions in a given node 
is based on the past wins and losses during previous simulations; this is what we call
a {\it bandit method}. Such a method, EXP3, is described in the next section.
One of the strengths of MCTS algorithms is that the tree $T_1$ which is built is asymmetric: some  branches
of $T$, consisting of nearly-optimal actions for both players, will be explored repeatedly,
but in the long run the whole tree $T$ will be explored.

A difficulty for the adaptation of these algorithms to the partially observable case is that when a player has to choose
his next action, he has to guess someway the unknown moves of his opponent. A standard method is to use a probability
distribution on the different possibilities of the opponent's past moves in order to estimate what will happen if an action is selected. This is
what we call {\it belief sampling}, and it has led to several implementations, using MCTS in 
a tree where only one player has choices and the opponent moves are predicted by different belief sampling methods
\cite{borsboom07,ciancarini09,parker05}.

These algorithms compute efficient strategies, but they are not intended to compute {\it solutions} of
the game, i.e. almost optimal strategies and Nash Equilibria, which is here our goal. 

On the other hand, a method named {\it minimization of counterfactual regret} has been
introduced in \cite{zink08} to compute Nash equilibria for partially observable games. However, as opposed to MCTS algorithms, this method has for each round
of computation to process the whole tree of the game, which is very long in most cases.

We propose here an alternative method which is aimed at computing Nash equilibria using MCTS algorithms. The method
has the main advantages of MCTS algorithms: it is consistent in the long run (convergence to a Nash Equilibrium)
but still efficient in the short term (asymmetry of the tree).

For the sake of conciseness we cannot develop further these notions apart from the specific algorithms that we use and refer to \cite{coulom06,kocsis06} for a general introduction to Monte-Carlo Tree Search and Upper Confidence Trees, and to \cite{audibert09,cesa06} for bandit methods.

\section{The EXP3 Algorithm }

This algorithm has been introduced in \cite{auer03} ; additional information can be found in \cite{audibert09}. We have the following framework:

\begin{enumerate}
  \item At each time-step $t \geq 0$, the player chooses a probability distribution on actions $\{1,2,\cdots,k\}$ ;
  \item Informed of the distribution, the environment secretly chooses a reward vector $(r^t_1,\cdots,r^t_k)$ ;
  \item An action $I_t \in \{1,\cdots,k\}$ is randomly chosen accordingly to the player's distribution, who then earns a reward $r^t_{I_t}$.
\end{enumerate}

The algorithm requires two parameters, $\gamma \in [0;1]$ and $\eta \in (O;\frac{1}{k}]$, which have to be tuned
(more informations in \cite{audibert09}). Both parameters control the ratio
between exploitation of empirically good actions and the exploration of insufficiently tested actions. If
one uses the algorithm with an infinite horizon, both parameters have to decrease to $0$.

\renewcommand{\algorithmiccomment}[1]{#1}
\begin{algorithm}
  \caption{EXP3 Algorithm} 
  \begin{algorithmic}[1]
    \STATE let $p^1$ be the uniform distribution on $\{1,\cdots,k\}$
    \FOR{ each round $t=1,2,\cdots$}
    \STATE choose randomly an action $I_t$ according to $p_t$ ; 
    \STATE update the expected cumulative reward of $I_t$ by
    $$G_{I_t} = G_{I_t} + \frac{r^t_{I_t}}{p^t_{I_t}}$$
    \STATE update the probability $p$ by setting for each $i \in \{1,\cdots,k\}$
    $$p^{t+1}_i = (1-\gamma) \frac{ \exp( \eta G_i)}{\sum_{j=1}^k \exp(\eta G_j)} + \gamma$$
    \ENDFOR
  \end{algorithmic} \label{exp3} 
\end{algorithm}

It can be proved that in a zero-sum matrix game - which is defined by a matrix $A$, where players respectively
choose a row $i$ and a column $j$ by a distribution probability, and where $A_{i,j}$ is the corresponding
reward for the first player (the other player earning the opposite) - if both players update
their probability distributions with the EXP3 algorithm, then the empirical distributions
of the players' choices converge almost surely to a Nash equilibrium.

\section{Our algorithm: Multiple Monte-Carlo Tree Search with EXP3 as a bandit tool}

We consider here partially observable games in extensive form, which does not necessarily mean that a tree is given, but rather are the rules of the game. More precisely, we suppose the existence of a {\it referee} able to compute, given the moves of each player, what is the new (secret) state of the game, and then sends observations to the players.

All players will separately run a MCTS algorithm, growing a tree depending on the other players' strategies; thus the whole algorithm behaves similarly to fictitious play \cite{brown51}. The nodes of these trees correspond to the successive interactions between players and the referee: moves of the player and observations. 
For each new simulation (i.e. single game) a new node is added to the tree for each player; during a game if a player has to move to a node which has not been constructed yet, then he stores information about this node and from this point plays randomly until the end of this game. At the end of the game, the node is added and results of this game are processed from this node up to the root of the tree.

We suppose for our implementation that the players have two different playing modes:

\begin{itemize}
  \item in {\it tree mode}, the player has in memory a {\it current node} corresponding to its history of successive moves and observations during the play. Each of these nodes have transitions corresponding to observations or moves, either leading to another existing node or leaving the tree if such a transition has never been considered. Players actualize their {\it current node} given the successive moves and observations, and if a transition leaves the tree then the player mode is set to {\it out of the tree}.
  \item in {\it out of tree mode}, players just play randomly with uniform probability on all moves.

    When a player is first set to {\it out of the tree mode}, a new node corresponding to the simulation is added, which we indicate in the algorithm by {\it first node out of the tree}. 

\end{itemize}

\renewcommand{\algorithmiccomment}[1]{#1}
\begin{algorithm}
  \caption{Multiple Monte-Carlo Tree Search with EXP3 policy for a Game in Extensive Form}
  \begin{algorithmic}[1]
    \REQUIRE a game G in extensive form
    \WHILE{(timeleft$>$0) }
    \STATE set players to tree mode and their current node to the roots of their trees 
    \REPEAT
    \STATE determine active player $i$\\
    \COMMENT get Player $i$'s move:\\
    \IF{Player $i$ is {\it tree mode}}
    \STATE choose randomly a move proportionally to the probabilities 
    $$p^i_m(N)=(1-\gamma(n))\frac{\mbox{rew}(N,m)}{\sum_{\ell=1}^{k(N)} \mbox{rew}(N,\ell)} +\frac{\gamma(n)}{k(N)}$$
    defined for all moves $m=1,\cdots k(N)$ from Player $i$'s current node $N$.
    \ELSE
    \STATE choose randomly the next move with uniform probability.
    \ENDIF
    \STATE return to all players {\it observations} according to the previous move.
    \FOR{each player $j$ in tree mode}
    \STATE determine the node $N'$ following the current node according to the observation.
    \IF{node $N'$ exists in memory}
    \STATE let $N'$ be the new current node of Player $j$
    \STATE store the probability $p(N')$ of the transition from $N$ to $N'$
    \ELSE
    \STATE store node $N'$ as the {\it first node out of the tree}
    \STATE set Player $j$ in {\it out of tree mode}
    \ENDIF
    \ENDFOR
    \UNTIL{game over} 
    \FOR{each player $j$}
    \STATE let $r_j$ be the reward obtained during the last play
    \IF{Player $j$ is in {\it out of tree mode}}
    \STATE add to Player $j$'s tree the {\it first node out of the tree}
    \STATE let $N$ be this node 
    \ELSE
    \STATE let $N$ be the last node encountered during the last play
    \ENDIF 
    \WHILE{$N \neq NULL$}
    \STATE update the reward of node $N$ for the move $m$ which was chosen in this node
    $$\mbox{rew}(N,m) \leftarrow \mbox{rew}(N,m) \cdot \exp\left(f(d)\frac{r_j}{p(N)}\right)$$
    where $d$ is the depth of node $N$ and $p(N)$ is the probability of the transition that led
    to node $N$ during the last play.
    \STATE do $N \leftarrow \mbox{father}(N)$
    \ENDWHILE
    \ENDFOR
    \ENDWHILE
  \end{algorithmic} \label{algo}
\end{algorithm}

Algorithm MMCTS requires two parameters that we now describe:

\begin{itemize}
  \item a function $\gamma$, depending on the number of simulations $n$, which is a parameter of the EXP3 algorithm used for mixing the exponentially weighted strategy with an uniform distribution. It is mandatory to have $\gamma$ tend to zero as the number $n$ of simulations goes to infinity, otherwise the empirical frequencies would remain close to a uniform distribution. Experimentally we used $\gamma(n)=n^{-0.3}$ in the case of phantom tic-tac-toe.
  \item a function $f$, depending on the depth $d$ of the nodes. This function is used to reward much more a node of great depth than a node close to the root for a good moves; the idea is that the success of a deep node is decisive, whereas a node close to the root leads to a lot of different strategies and we should be careful by not rewarding it to much for single success. We used $f(d)=1.7^{d-9}$.
\end{itemize}

Clearly these parameters have to be tuned and our choices are empirical.

\section{Experimental results}

We test our algorithm in the simple context of phantom tic-tac-toe. While being simpler than other phantom games, the full tree of possible moves for a single player is quite huge. Whereas the classic tic-tac-toe game is totally deterministic, and known to end up with a draw if both players play optimally, in the phantom case the partial observability leads the players to consider mixed strategies for their moves. Thus it will not be surprising that if both players play optimally, with a little luck both can win a single game. If both player play totally randomly with uniform probabilities (which applies as well to the classic and phantom settings), Player 1 wins 60 \% of the matches and Player 2\% about 30\% (thus 10\% are draws) - see Table \ref{tableMatches}; thus clearly the game favors Player 1. The strategy stealing argument shows that this is also the case in the phantom case if both players play optimally. 
What is more surprising is that we obtain:\\

{\noindent\bf Experimental result} The value of the game is approximatively 0.81.\\

\begin{figure} 
  \includegraphics[width=11cm]{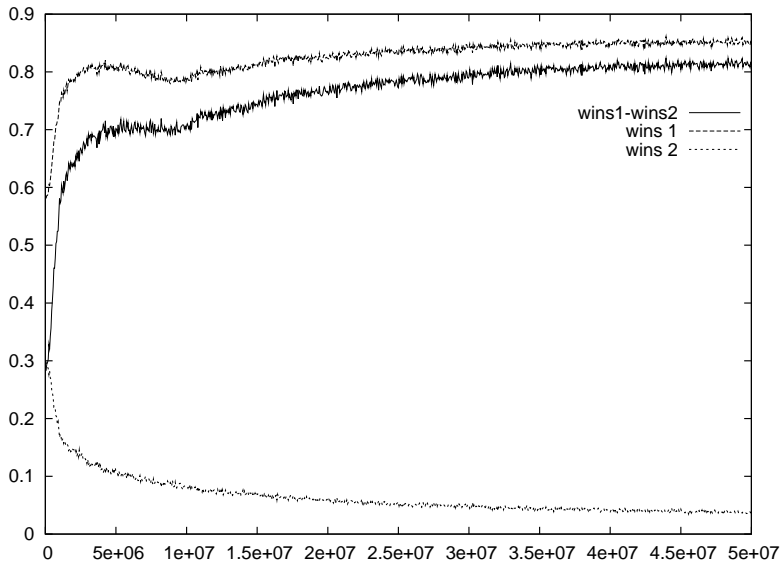}
  \caption{Probabilities of winning for Player 1, Player 2 and their difference according to the number of simulations. The difference
  converges to the value of the game 0.81 .}\label{1vs2}
\end{figure}

We refer to classic textbooks in Game Theory (e.g. \cite{fudenberg91}) for the definition of value of a zero-sum game or Nash equilibrium. Here the value is to be understood with a score of $+1$ if Player 1 wins and a score of $-1$ if Player 2 wins (and $0$ for a draw). Figure \ref{1vs2} depicts the evolution of the number of wins of Player 1 and Player 2 as the number of simulations grows.

In fact Player 1 can force about 85 \% of victory whereas Player 2 can force only about 4 \% of victory. 
We now present some competitors that we designed to test our algorithm. The results of repeated matches betweens
these players are given on Table \ref{tableMatches}.\\
 
{\noindent \bf The Random Player:} plays every move randomly with uniform probability.\\

{\noindent \bf The Belief Sampler Player:} this player uses belief sampling as described in the introduction.
He has in memory the full tree of classic observable tic-tac-toe, and before each move considers all the possible sets of moves
of the opponent that match the current state of observations, and stores optimal moves. It then randomly decides a move  proportionally to the frequencies obtained
during the previous simulation. This is a quite strong opponent: see the results of the matches opposing Belief Sampler and Random Player 
on Table \ref{tableMatches}. However, the results of matches Belief Sampler versus Belief Sampler are far from the value of the game, and
are exactly the same that we obtain if both players play at random (Table \ref{tableMatches}).\\

{\noindent \bf The MMCTS Players:} these are the players that we obtain after letting algorithm MMCTS run for a given number of simulations. We chose these numbers to be 500,000, 5 millions and 50 million simulations.
Observe that as a first player, only Belief Sampler can stand the pressure against MMCTS 50M 
but as a second Player only the former resists against all opponents. For instance, it appears that Belief Sampler is a better Player 2 against Random Player than MMCTS 50M is, however MMCTS 50M always ensures a good proportion of wins. Also observe that in MMCTS 50M versus Belief Sampler matches, our player is much better.

\begin{table}
  \renewcommand\arraystretch{1.4}
  \scalebox{.9}{
  \begin{footnotesize}   
    \begin{tabular*}{1.1\textwidth}{@{\extracolsep{\fill}}|c||c|c|c||c|c|}
      \hline
      Player 1 \tb \ Player2 \ &   MMCTS 500K  &   MMCTS 5M  &   MMCTS 50M  &  Random   &  Belief Sampler   \\
      \hline
      \hline
      MMCTS 500K                        & $65 \%$ \tb \ $ 25 \%$ & $51 \%$ \tb \ $ 37 \%$ & $44 \%$ \tb \ $ 47 \%$ & $67 \%$ \tb \ $ 22 \%$ & $40 \%$ \tb \ $ 43 \%$\\
      \hline
      MMCTS 5M                          & $88 \%$ \tb \ $ 06 \%$ & $82 \%$ \tb \ $ 10 \%$ & $78 \%$ \tb \ $ 17 \%$ & $88 \%$ \tb \ $ 05 \%$ & $78 \%$ \tb \ $ 10 \%$\\
      \hline
      MMCTS 50M                         & $93 \%$ \tb \ $ 02 \%$ & $89 \%$ \tb \ $ 03 \%$ & $85 \%$ \tb \ $ 04 \%$ & $93 \%$ \tb \ $ 02 \%$ & $82 \%$ \tb \ $ 03 \%$\\
      \hline
      \hline
      Random                            & $55 \%$ \tb \ $ 33 \%$ & $48 \%$ \tb \ $ 39 \%$&$41 \%$ \tb \ $ 47 \%$ & $59 \%$ \tb \ $ 28 \%$ & $30 \%$ \tb \ $ 53 \%$\\
      \hline
      Belief Sampler                    & $77 \%$ \tb \ $ 14 \%$ & $73 \%$ \tb \ $ 18 \%$ & $68 \%$ \tb \ $ 22 \%$ & $79 \%$ \tb \ $ 12 \%$ & $56 \%$ \tb \ $ 28 \%$ \\
      \hline
    \end{tabular*} 

  \end{footnotesize}
  }
  \vspace{.1cm}
  \caption{Probability of winning a game for Player 1 \tb Player2.}
  \label{tableMatches}
\end{table}

\vspace{-.5cm}
Let us explain now why we pretend that the strategies of the MMCTS 50M players are ``approximatively optimal strategies''. By approximatively optimal, we mean that the strategy behaves like a Nash equilibrium strategy - it ensures a certain value - versus most opponent strategies. 
In order to compute really optimal strategies, one would have to let the algorithm run for a very long time. However, even with 50 Million simulations (which takes less than an hour on a standard computer) the asymmetric trees that have been grown contain most of the branches corresponding to high probability moves in a real Nash equilibrium. Nevertheless, in the short term these strategies cannot be perfect, and branches less explored can be used
by opponents to build a strategy specifically designed to beat our algorithm.

\begin{figure}[h!] 
  \includegraphics[width=11cm]{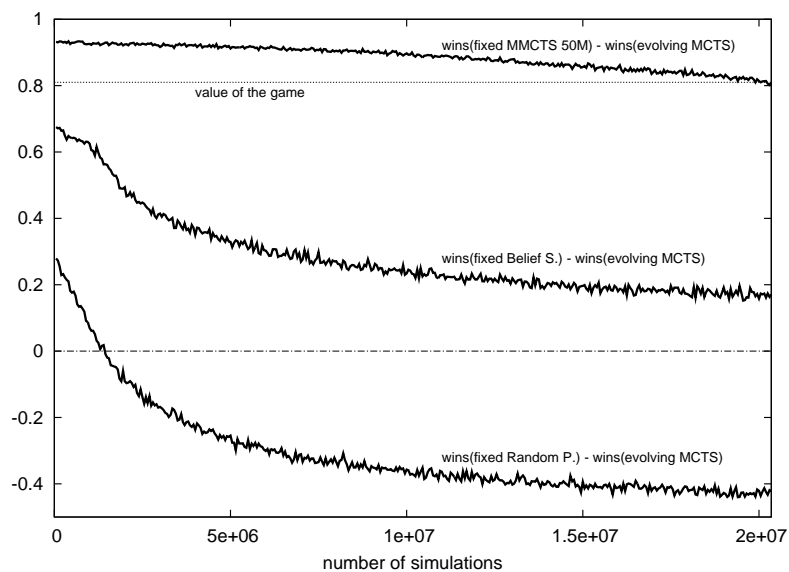}
  \caption{Performance of the fixed players MMCTS 50M, Belief Sampler and Random Player (as first players) against an opponent evolving by a simple MCTS: difference of the probabilities
  of winning a single game for Player 1 and Player 2.}
  \label{nashitude1}
\end{figure}

\begin{figure}[h!] 
  \includegraphics[width=11cm]{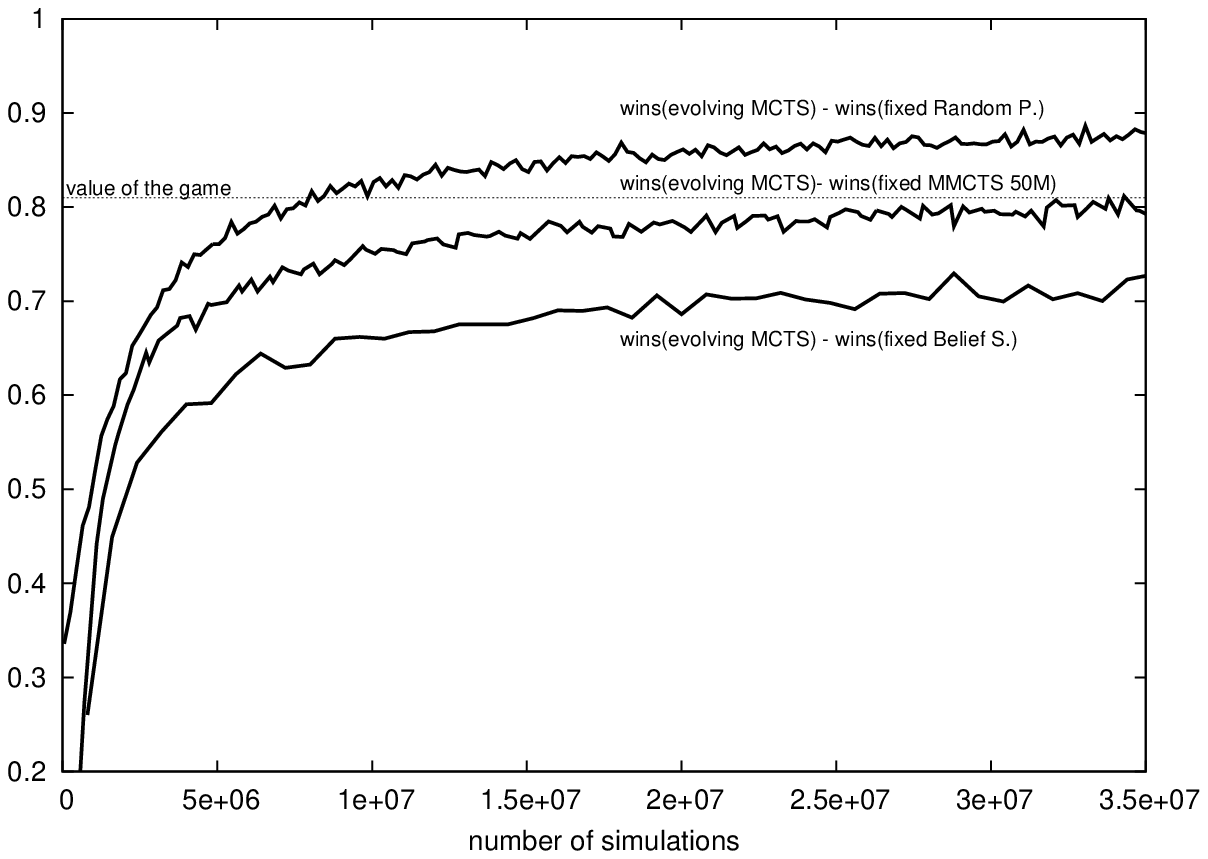}
  \caption{Performance of the fixed players MMCTS 50M, Belief Sampler and Random Player (as second players) against an opponent evolving by a simple MCTS: difference of the probabilities
  of winning a single game for Player 1 and Player 2.}
  \label{nashitude2}
\end{figure}

A way to test this is to fix the strategies obtained by our algorithm and to have them compete with an opponent initialized as a random player and evolving with a one-sided MCTS. At last the evolving opponent will be able to spot weaknesses and exploit them. Hence a way to measure a player's robustness is to test whether he can stand in the long run when opposed to an evolving opponent.
We depict on Figures \ref{nashitude1} and \ref{nashitude2} the evolutions of the difference of wins for Random Player, MMCTS 50M and Belief Sampler against an evolving opponent, which is respectively the second and the first player on Fig. \ref{nashitude1} and Fig. \ref{nashitude2}.

We observe that as a first player (Fig. \ref{nashitude1}), MMCTS 50M resists in the long run to all attacks from the evolving opponent, whereas Random Player and Belief Sampler are defeated way below the value of the game (of course if we wait much longer it will also be the case for MMCTS 50M); here the supremacy of MMCTS 50M is undeniable. As a second player (Fig. \ref{nashitude2}) its performance is less spectacular and Belief Sampler seems to resist much better to the assaults of the evolving opponent; however MMCTS does what it is built for, {\it i.e.} ensure the value of the game regardless of the opponent.

\section{Conclusion}

In this paper we showed a way to adapt Monte-Carlo tree search algorithms to the partially observable case in order to compute Nash equilibria of these games. We proposed the MMCTS algorithm, which we used as an experimental example  in the case of phantom tic-tac-toe, obtaining strong players and the approximative value of the game. In particular, the strength of our player was proved by its resistance when fixed against an evolving player, and its good results against one of the best players known for partially observable games, the Belief Sampler Player.
The experimental results being promising, we have several directions for future research. First, we must obtain bounds on the convergence of the algorithm to a Nash equilibrium, and find a way to rigorously define the notion of ``very good versus most strategies'' that we described and tested. Second, it will be necessary
to implement the algorithm in a larger framework, for instance for kriegspiel or poker. Finally, a problem still open is to how compute optimal strategies
with MCTS algorithms without starting from the root of the tree but from any observed position: 
this seems to involve necessarily beliefs on the real state of the game. How can one compute  these beliefs without starting from the root ? Progress has to be made
with MCTS algorithms before solving this question.

\addtolength{\extrarowheight}{15pt}

\bibliographystyle{plain}
\bibliography{auger}

\begin{thebibliography}{10}

\bibitem{audibert09}
J.Y. Audibert and S.~Bubeck.
\newblock {Minimax policies for adversarial and stochastic bandits}.
\newblock In {\em Proceedings of the 22nd Annual Conference on Learning Theory,
  Omnipress}, 2009.

\bibitem{auer03}
P.~Auer, N.~Cesa-Bianchi, Y.~Freund, and R.E. Schapire.
\newblock {The nonstochastic multiarmed bandit problem}.
\newblock {\em SIAM Journal on Computing}, 32(1):48--77, 2003.

\bibitem{borsboom07}
J.~Borsboom, J.~Saito, G.~Chaslot, and J.~Uiterwijk.
\newblock {A comparison of Monte-Carlo methods for Phantom Go}.
\newblock In {\em Proc. 19th Belgian--Dutch Conference on Artificial
  Intelligence--BNAIC, Utrecht, The Netherlands}, 2007.

\bibitem{brown51}
G.W. Brown.
\newblock {Iterative solution of games by fictitious play}.
\newblock {\em Activity analysis of production and allocation}, 13(1):374--376,
  1951.

\bibitem{cazenave06}
T.~Cazenave.
\newblock {A Phantom-Go program}.
\newblock {\em Advances in Computer Games}, pages 120--125, 2006.

\bibitem{cesa06}
N.~Cesa-Bianchi and G.~Lugosi.
\newblock {\em {Prediction, learning, and games}}.
\newblock Cambridge Univ Pr, 2006.

\bibitem{ciancarini09}
P.~Ciancarini and G.P. Favini.
\newblock {Monte carlo tree search techniques in the game of Kriegspiel}.
\newblock In {\em Proceedings of the Twenty-First International Joint
  Conference on Artificial Intelligence (IJCAI-09)}, pages 474--479, 2009.

\bibitem{coulom06}
R.~Coulom.
\newblock {Efficient selectivity and backup operators in Monte-Carlo tree
  search}.
\newblock In {\em Proceedings of the 5th international conference on computers
  and games}, pages 72--83, 2006.

\bibitem{fudenberg91}
D.~Fudenberg and J.~Tirole.
\newblock {\em {Game Theory}}.
\newblock MIT Press, 1991.

\bibitem{kocsis06}
L.~Kocsis and C.~Szepesv{\'a}ri.
\newblock {Bandit based monte-carlo planning}.
\newblock {\em Machine Learning: ECML 2006}, pages 282--293, 2006.

\bibitem{li94}
D.H. Li.
\newblock {\em {Kriegspiel: Chess Under Uncertainty}}.
\newblock Premier Pub. Co., 1994.

\bibitem{parker05}
A.~Parker, D.~Nau, and VS~Subrahmanian.
\newblock {Game-tree search with combinatorially large belief states}.
\newblock In {\em International Joint Conference on Artificial Intelligence},
  volume~19, page 254, 2005.

\bibitem{zink08}
M.~Zinkevich, M.~Johanson, M.~Bowling, and C.~Piccione.
\newblock {Regret minimization in games with incomplete information}.
\newblock {\em Advances in Neural Information Processing Systems},
  20:1729--1736, 2008.

\end{thebibliography}

\end{document}